\documentclass[aps,prc,reprint,superscriptaddress,nofootinbib]{revtex4-2}

\usepackage{amsmath,amssymb,amsfonts}
\usepackage{graphicx}
\usepackage{hyperref}
\usepackage{cleveref}
\usepackage{bm} 
\usepackage{physics} 
\usepackage{xcolor}

\newcommand{\He}[1]{$^{#1}$He}
\newcommand{\HA}{\mathcal H}
\newcommand{\Ht}{\tilde{\mathcal H}}
\newcommand{\RE}{\mathfrak Re\ }
\newcommand{\IM}{\mathfrak Im\ }

\begin{document}

\title{Polarized Nuclear DVCS at the EIC}

\author{J. R. Pybus}
\affiliation{Los Alamos National Laboratory, Physics Division, Los Alamos, NM, 87545, USA}
\author{X. Li}
\affiliation{Los Alamos National Laboratory, Physics Division, Los Alamos, NM, 87545, USA}
\author{L. Calero Diaz}
\affiliation{Los Alamos National Laboratory, Physics Division, Los Alamos, NM, 87545, USA}
\date{\today}

\begin{abstract}
The Electron-Ion Collider (EIC) will enable a series of measurements at unprecedented energies and luminosities, providing new opportunities to investigate the microscopic structure of nucleons and nuclei at small $x_B$. Exclusive processes such as Deeply Virtual Compton Scattering (DVCS) offer unique access to the three-dimensional structure of hadrons through Generalized Parton Distributions (GPDs), while polarized electron and ion beams further enable detailed studies of spin-dependent structure.
A model for coherent DVCS on polarized \He3 is developed and applied to simulations of $9\times166$-GeV/nucleon $e$\He3 collisions at the EIC. Using this framework, the statistical precision achievable is estimated for measurements of beam-spin asymmetries and for the extraction of the Compton Form Factors (CFFs) $\HA_{^3\mathrm{He}}$ and $\Ht_{^3\mathrm{He}}$.
Early EIC data are found to enable precise differential measurements of the unpolarized CFF $\mathcal H_{^3\mathrm{He}}$ and to provide significant constraints on its real and imaginary components. By contrast, meaningful constraints on the polarized CFF $\Ht_{^3\mathrm{He}}$ require substantially larger integrated luminosities. The kinematics of the recoil \He3 nucleus are also examined, and the advanced far-forward detector capabilities at the EIC required to tag the intact nucleus and perform fully exclusive measurements of coherent nuclear DVCS are discussed.
\end{abstract}

\maketitle

\section{Introduction}

An important aim of nuclear physics in the modern era is to advance our understanding of the three-dimensional partonic structure of hadrons, encoded within the framework of Generalized Parton Distributions (GPDs)~\cite{Muller1994,Ji1_97,Ji2_97,Rad_96,Rad_97,Diehl_2003}, which carry both momentum and spatial information about the quarks and gluons within a hadronic state.
GPDs can be accessed via a number of exclusive scattering processes, the most commonly used of which is Deeply Virtual Compton Scattering (DVCS).
DVCS is the hard exclusive electroproduction of a real photon from a hadronic target and represents a clean probe of GPDs. 
Many measurements of this process have been performed and are ongoing at 
HERA~\cite{
Airapetian2001,
Aktas2005,
Airapetian2007,
Airapetian2008,
Aaron2009,
Airapetian2009,
Chekanov2009,
Airapetian2010,
Airapetian2011,
Airapetian2012}, Jefferson Lab~\cite{
Stepanyan2001,
Camacho2006,
CLAS06_Pol,
PhysRevLett.99.242501,
CLAS08_Asym,
CLAS09_Asym,
JoCLAS:2015,
CLAS15_Asym,
Defurne2015,
Defurne2017,
HallANaturePhys2020,
Georges2022}, and COMPASS~\cite{
Akhunzyanov2019,
Akhunzyanov2019err,
PhysRevLett.130.211902,
PhysRevLett.133.211903}.


The majority of existing DVCS measurements have been carried out on the proton and are aimed at accessing its GPDs.
Relatively few measurements exist to constrain the three-dimensional partonic structure of nuclei~\cite{PhysRevC.81.035202,Hattawy_2017,
CLAS:2021ovm}, and even fewer can reliably distinguish the coherent signal from the incoherent background. 
The partonic structure of nuclei remains an outstanding topic in nuclear physics~\cite{Aubert83,Arnold:1984,Ashman88,Arneodo90,Gomez94,Seely09,Hen:2016kwk,Schmookler:2019nvf}, and DVCS has the potential to broaden this topic into the area of tomography.

\begin{figure}
    \centering
    \includegraphics[width=1\linewidth]{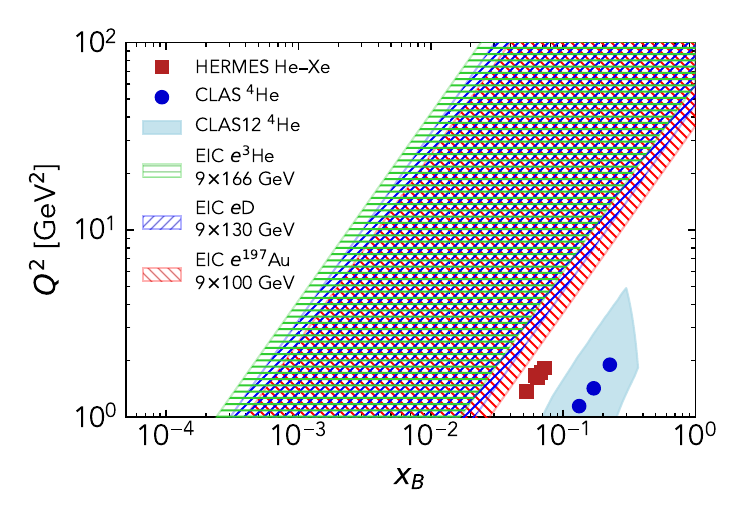}
    \caption{Kinematic coverage in $x_B$ and $Q^2$ of coherent nuclear DVCS measurements. 
    Red squares and blue circles show the coverage of measurements from HERMES~\cite{PhysRevC.81.035202} and CLAS~\cite{Hattawy_2017} respectively, while the blue shaded region shows the projected coverage from CLAS12 data~\cite{armstrong2017partonicstructurelightnuclei}. 
    The hatched regions show the kinematic coverage possible with different electron-ion systems at the EIC, including $e^3$He at $9$ GeV$\times166$ GeV/nucleon (green, horizontal hatch), $e$D at $9$ GeV$\times130$ GeV/nucleon (blue, forward-diagonal hatch), and $e^{197}$Au at $9$ GeV$\times100$ GeV/nucleon (red, backward-diagonal hatch), each requiring $0.01<y<0.7$.}
    \label{fig:coverage}
\end{figure}

Exclusive processes such as DVCS are an important component of the physics program of the future Electron-Ion Collider (EIC)~\cite{Abdul_Khalek_2022,Bylinkin_2023,Adkins:2022jfp, ATHENA:2022hxb, fy8y-bjc9,boer2025}, which aims to constrain the three-dimensional structure of the proton and nuclei across a wide kinematic range in $x_B$ and $Q^2$, as shown in Fig.~\ref{fig:coverage} for nuclear DVCS measurements, accessing qualitatively different parton dynamics than current data sets~\cite{PhysRevC.81.035202,Hattawy_2017} and ongoing experiments~\cite{armstrong2017partonicstructurelightnuclei}.
The EIC will utilize a series of polarized and unpolarized electron-proton ($ep$) and electron-nucleus ($e$A) collisions to access higher center-of-mass energies (20--141~GeV) \cite{Abdul_Khalek_2022} than those available in fixed-target experiments. 
The current design of the EIC project detector, ePIC, can precisely determine the kinematics of key observables, such as the scattered electrons and the produced real photon in the DVCS process.
The ePIC far-forward region detectors, which cover small angles $\theta\lesssim 25$ mrad in the forward hadron-going direction \cite{Abdul_Khalek_2022}, are crucial for performing exclusive measurements by tagging particles experiencing very small momentum-transfer.
This small-angle tagging is also a key capability enabling the EIC to perform exclusive measurements of nuclei. 
By measuring nucleons and nuclear fragments from the dissociation of beam nucleus, coherent and incoherent nuclear reactions may be distinguished.
Such tomographic measurements of the nucleus represent key constraints on the three-dimensional distribution of the partons within, shedding light on how the nuclear medium impacts the behavior of partons beyond the one-dimensional $x$-dependent modification probed in Deep-Inelastic Scattering and Drell-Yan measurements~\cite{Kovarik:2015cma,Eskola:2021nhw,Helenius:2021tof,Segarra:2020gtj,AbdulKhalek:2022fyi,Ruiz:2023ozv,nCTEQ:2023cpo}.

\section{Formalism and Model}

\begin{figure}
    \centering
    \includegraphics[width=1\linewidth]{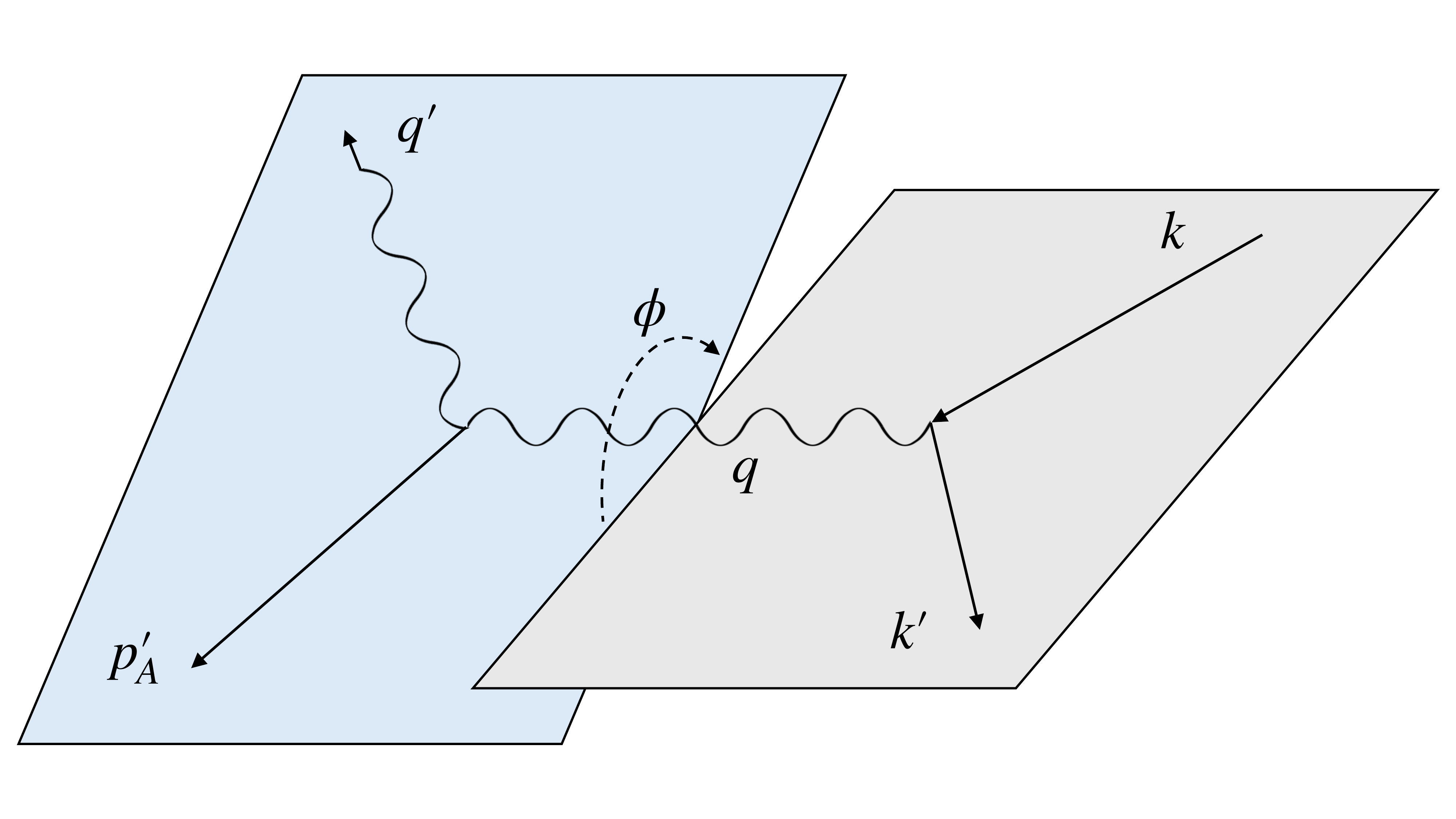}
    \caption{Kinematics of the coherent DVCS process in the initial-state ion rest frame. The azimuthal angle $\phi$ is defined between the lepton and hadron planes using the Trento convention.}
    \label{fig:DVCS_kinematics}
\end{figure}

The theoretical formalism relating the DVCS cross section to the target GPDs has been extensively developed~\cite{Ji2_97,Vanderhaeghen_1999,Belitsky_2002,Boffi_2008}, including its extension to nuclear targets ~\cite{Kirchner_2003,PhysRevC.68.015204,Liuti_2005,Guzey_2008,Fucini_2018,Fucini_2021,martínezfernández2026coherentdeeplyvirtualcompton,martínezfernández2026quarkgluontomographyhelium4}.
In this study, we follow the leading-twist formulations of Ref.~\cite{Belitsky_2001} and Ref.~\cite{Belitsky_2002} for spin-0 and spin-1/2 hadrons, respectively, while neglecting contributions from gluon transversity.
The total cross section for the exclusive leptoproduction of a photon from nucleus $A$, $lA\rightarrow l\gamma A$, is given by
\begin{equation}
    \frac{d\sigma}{dx_Adydtd\phi}=\frac{\alpha^3x_Ay}{8\pi Q^2\sqrt{1+\epsilon^2}}\left|\frac{\mathcal{T}}{e}\right|^2\,.
\end{equation}
The lepton has initial and final four-momentum $k$ and $k'$, while the nucleus has initial and final four-momentum $p_A$ and $p_A'$, respectively. We define the lepton squared momentum-transfer $Q^2=-q^2=-(k-k')^2$, the hadron squared momentum-transfer $t=\Delta^2=(p_A'-p_A)^2<0$, the nuclear Bjorken variable $x_A=Q^2/2p_A\cdot q\approx x_B/A$, the lepton energy-fraction $y=q\cdot p_A/k\cdot p_A$, and the target-mass kinematic factor $\epsilon=2x_A m_A/Q$, along with the azimuthal angle $\phi$ between the lepton and hadron scattering planes in the nuclear rest frame, where we follow the Trento convention~\cite{PhysRevD.70.117504} in the following.
Also relevant is the ``skewness'' $\xi_A\approx\frac{x_A}{2-x_A}$, which characterizes the longitudinal momentum-transfer to the nucleus.
Fig.~\ref{fig:DVCS_kinematics} illustrates the kinematics of the reaction as well as the convention for $\phi$ used here.

A key feature of the DVCS process is its irreducible interference with the purely QED Bethe Heitler (BH) process, in which the real photon is emitted by either the incoming or outgoing lepton. 
As a result, the squared amplitude for the process can be written as
\begin{equation}
    |\mathcal T|^2=|\mathcal T_{BH}|^2+|\mathcal T_{DVCS}|^2+\mathcal I ,
\end{equation}
where DVCS-BH interference term is defined by
\begin{equation}
    \mathcal I = \mathcal T_{DVCS}\mathcal T_{BH}^* +\mathcal T_{BH}\mathcal T_{DVCS}^*\,.
\end{equation}
In practice, separating the contribution from different CFFs is most readily achievable by examining the DVCS-BH interference terms, which can be isolated via polarization and/or lepton-charge asymmetries.
These asymmetries also prove favorable for experimental interpretation, reducing systematic effects and the need for absolute normalization of data.

More generally, the BH, DVCS, and interference contributions to the cross section can be decomposed into Fourier harmonics of the hadronic azimuthal angle $\phi$:

\begin{equation}
\begin{aligned}
|\mathcal T_{DVCS}|^2,&\;|\mathcal T_{BH}|^2,\;\mathcal I
= \\ 
&\mathcal K \Bigg[
c_0
+ \sum_{n=1}^2
\left(
c_n \cos(n\phi)
+ s_n \sin(n\phi)
\right)
\Bigg]\,,
\end{aligned}
\end{equation}
where the prefactors $\mathcal K$ and the harmonic coefficients $c_n$ and $s_n$ differ for BH, DVCS, and the interference term.
The harmonic coefficients have different expressions for spin-0 and spin-1/2, given by Ref.~\cite{Belitsky_2001} and Ref.~\cite{Belitsky_2002} respectively, neglecting higher-order kinematical effects~\cite{PhysRevD.82.074010} which are suppressed in the high-energy kinematics of the EIC.
The BH coefficients depend bilinearly on the electromagnetic form factors of the hadron, while the DVCS coefficients depend bilinearly on the ``Compton Form Factors'' (CFFs), which are related to sums over the individual quark GPDs within a target weighted by the square of the quark charge.
The interference coefficients are determined by products of electromagnetic and Compton form factors. 
In general, these coefficients may or may not depend also on the lepton helicity $\lambda=\pm 1$, and include different terms to account for the hadron polarization vector $\Lambda$:
\begin{equation}
    d\sigma=d \sigma_\text{unp} + \Lambda \cos \theta_{\Lambda} d\sigma_{LP} + \Lambda \sin \theta_\Lambda d\sigma_{TP}(\phi_{\Lambda})\,,
\end{equation}
where $d\sigma_\text{unp}$, $d\sigma_{LP}$, and $d\sigma_{TP}$ denote the cross section for unpolarized, longitudinally-polarized, and transversely-polarized hadrons, respectively, and each include different harmonic coefficients.

In this study, we focus on the nuclei \He4 and \He3, which have spin 0 and spin $1/2$, respectively. 
For \He4, we employ the charge form factor calculated in Ref.~\cite{PhysRevC.43.1585}. For \He3, we use the electric and magnetic form factors obtained from the phenomenological fits of Ref.~\cite{Barcus:2019igm}.

Nuclear CFFs are considerably less constrained than their nucleonic counterparts, with only limited experimental data and theoretical calculations currently available. 
In this work, we evaluated the nuclear CFFs with the impulse approximation, expressing them as convolutions of the corresponding nucleon CFFs as detailed in Refs.~\cite{Scopetta_2004,Scopetta_2009,Rinaldi_2013}. 
At leading twist, a spin-1/2 nucleus is characterized by four chirality-conserving CFFs (as well as four additional CFFs associated with gluon transversity). 
The dominant contribution, which is also the only chirality-conserving leading-twist CFF for a spin-0 nucleus, is the unpolarized helicity-conserving CFF $\mathcal H$.
For both \He3 and \He4, the nuclear CFF $\mathcal H_A$ can be related to the nucleon CFFs $\mathcal H_N$ through the convolution formalism~\cite{Scopetta_2004,Scopetta_2009,Rinaldi_2013}:
\begin{equation}
\mathcal H_A(\xi_A,t) = \sum_N \int_\xi^1\frac{dz}{z} h^A_N(z,\xi_A,t) \mathcal H_N\left(\frac{\xi_A}{z},t\right)\,,
\end{equation}
where $h^A_N(z,\xi_A,t)$ is the off-diagonal light-cone momentum distribution for nucleon $N$ in nucleus $A$. 
For small values of the momentum transfer square $t$ and skewness $\xi_A$, the off-forward light-cone momentum distribution factorizes approximately as:
\begin{equation}
    h^A_N(z,\xi_A,t) \approx h_N^A(z)\times F_N^A(t)\,.
\end{equation} 
Here, $h_N^A(z)$ is the light-cone momentum distribution of nucleon $N$ in nucleus $A$, with momentum fraction $z\approx p^+/p_A^+$, while $F_N^A(t)$ characterizes the spatial distribution of nucleon $N$ through its contribution to the nuclear form factor.

For \He3 and other spin-$1/2$ nuclei, an additional leading-twist CFF, the polarized helicity-conserving CFF $\tilde {\mathcal H}$, contributes. Within the impulse approximation, this CFF can be expressed in terms of the corresponding nucleon CFFs:
\begin{equation}
\Ht_A(\xi_A,t) = \sum_N \int_{\xi_A}^1\frac{dz}{z} \tilde h^A_N(z,\xi_A,t) \tilde {\mathcal H}_N\left(\frac{\xi_A}{z},t\right)\,,
\end{equation}
where $\tilde h^A_N(z,\xi_A,t) = h^A_{N,\uparrow}(z,\xi_A,t) - h^A_{N,\downarrow}(z,\xi_A,t)$ is the net polarized off-diagonal light-cone momentum distribution.
In the absence of detailed spin-dependent nuclear structure calculations, we approximate the polarized off-forward distribution by multiplying the corresponding unpolarized distribution by the effective per-nucleon polarization $P_N^A$. For \He3, we adopt $P_n^{^3\text{He}}=0.86$ and $P_p^{^3\text{He}}=-0.028$~\cite{Bissey_2002}.
In addition, spin-$1/2$ nuclei are characterized by unpolarized and polarized helicity-flip CFFs $\mathcal E$ and $\tilde{\mathcal E}$. These quantities contribute to transverse-spin observables and are sensitive to more complex nuclear dynamics. Accordingly, these are set to 0 in the present analysis.

Combining the cross-section formalism outlined above (where only $\HA$ and $\Ht$ are considered) with appropriate models of nucleon and nuclear structure, we construct complete descriptions of coherent photon leptoproduction from spin-0 and spin-$1/2$ nuclei.
The resulting cross sections are implemented in a Monte Carlo event generator to produce pseudodata samples, which can be used both to benchmark the model against existing measurements and to evaluate projected rates and sensitivities for future experiments.
In the following sections, we first compare our calculations for \He4 with measurements from HERMES and CLAS experiments to assess the validity of the convolution approach employed in this study. We then apply the validated framework to predict coherent DVCS observables for polarized \He3 and to evaluate the corresponding experimental sensitivities at the Electron-Ion Collider (EIC) using the ePIC detector acceptance.

As input to the model, we employ the KM15 parametrization of the nucleon CFFs~\cite{Kumeri_ki_2010}, using the parameter values reported in Ref.~\cite{sb63-sfdt}.
The nuclear form factors $F_N^A(t)$ are obtained by Fourier transforming the coordinate-space nucleon densities of Ref.~\cite{PhysRevC.89.024305}; these are distinct from the electromagnetic nuclear form factors of Refs.~\cite{PhysRevC.43.1585,Barcus:2019igm}, as these include only the spatial distribution of nucleons inside the nucleus, and not the net charge or magnetic effects.
For \He3, the nonrelativistic spectral function of Ref.~\cite{Ciofi_degli_Atti_2005} is used to construct the associated light-cone momentum distribution. 
For nucleon momentum $\vec k$ in the nuclear rest-frame, the light-cone momentum fraction is approximated by 
\begin{equation}
    z \approx \frac{\sqrt{\vec k^2+m_N^2}+k_z}{\sqrt{\vec k^2+m_N^2}+\sqrt{\vec k^2+m_{A-1}^2}}\,,
\end{equation}
which exhibits the correct asymptotic behavior and remains well behaved in the limit of large $|\vec k|$.
For \He4, light-cone momentum density is derived from the momentum distribution of Ref.~\cite{PhysRevC.89.024305}, assuming that the $A-1$ system remains bound.
To account for nuclear shadowing effects, we further implement the $A$-dependent rescaling of $x_A$ introduced in Ref.~\cite{Vitev_2006} when evaluating the nuclear CFFs.

We emphasize that the process considered in this work is coherent DVCS, in which the nucleus remains intact in the final state. This exclusivity condition is essential for isolating the static off-forward components of the nucleus encoded in its generalized parton distributions. 
By contrast, the incoherent DVCS (and BH) process, $lA\rightarrow l\gamma X$, involves nucleus excitation or breakup and is therefore sensitive to nuclear fluctuations and the properties of bound nucleons. Consequently, coherent and incoherent photon-production processes must be distinguished experimentally.
Although incoherent scattering is not considered explicitly in the present study, it constitutes a potential background to coherent DVCS when complete final-state exclusivity cannot be enforced. This background becomes increasingly important with increasing $|t|$ and eventually dominates the total cross section at sufficiently large momentum transfer.

\section{Comparison with \He4 in Fixed-Target Experiments}

To validate both the convolution model described above and its implementation in the event generator, we compare our calculations for \He4 with existing measurements from CLAS~\cite{Hattawy_2017} and HERMES~\cite{PhysRevC.81.035202}.
In both experiments, information on the nuclear CFF $\mathcal H$ was accessed through the lepton single-spin asymmetry,
\begin{equation}
    A_{LU} = \frac{1}{P_l}\frac{d\sigma^\rightarrow - d\sigma^\leftarrow}{d\sigma^\rightarrow + d\sigma^\leftarrow}\,,
\end{equation}
where $d\sigma^{\rightarrow(\leftarrow)}$ denotes the differential cross section for positive (negative) lepton helicity, and $P_l$ is the degree of lepton polarization.
Changing the helicity of the lepton beam does not affect the BH or DVCS contributions to the total cross section, but it reverses the sign of the helicity-dependent harmonic coefficients in the interference term, most notably $s_1^{\mathcal I}$.
When the total cross section is dominated by the Bethe--Heitler contribution, the beam-spin asymmetry can be approximated by the following linearized expression:
\begin{equation}
    A_{LU}= \frac{x_A(1+\epsilon^2)^2}{y}\frac{s_1^{\mathcal I}\sin \phi+\cdots}{c_0^{BH}+\cdots}\propto \IM \mathcal H(\xi_A ,t)\sin \phi\,.\label{eq:ALU_lin}
\end{equation}
These measurements therefore focused on the $\sin \phi$ harmonic coefficient, $A_{LU}^{\sin \phi}$, in order to constrain the imaginary component of $\mathcal H^{^4\text{He}}(\xi_A ,t)$ as a function of the kinematic variables.
Throughout this study, we extract this coefficient by fitting the $\phi$-dependence of $A_{LU}$ within a given kinematic bin according to
\begin{equation}
    A_{LU}(\phi) = \frac{A_{LU}^{\sin \phi}\sin \phi}{1 + B \cos \phi}\,,
    \label{eq:fit}
\end{equation}
where higher-order harmonic contributions have been neglected, apart from the leading $\cos \phi$ modulation in the denominator. The latter arises primarily from the azimuthal dependence of the unpolarized cross section and retains some sensitivity to the real part of $\mathcal H$ ($\RE \mathcal H$).

\begin{figure}
    \centering
    \includegraphics[width=1\linewidth]{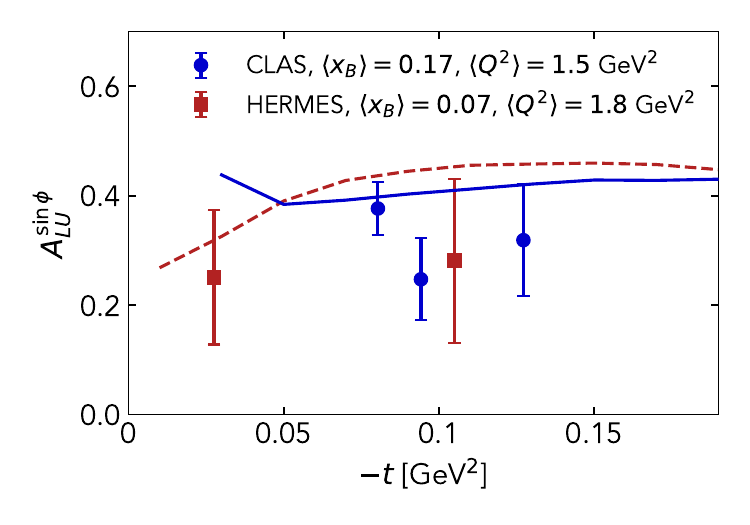}
    \caption{$A_{LU}^{\sin \phi}$ measurements from \He4 using 6-GeV electrons at CLAS~\cite{Hattawy_2017} (blue circles) and 27.6-GeV positrons at HERMES~\cite{PhysRevC.81.035202} (red squares), compared with calculations using matching kinematics (blue solid and red dashed lines, respectively), as a function of $-t$.}
    \label{fig:fixed-target}
\end{figure}

In Fig.~\ref{fig:fixed-target}, we compare the predictions of our model with the results reported in Refs.~\cite{Hattawy_2017,PhysRevC.81.035202}.
For the HERMES measurements, the event selection is not fully exclusive and therefore likely contains some contamination from incoherent scattering. Accordingly, we restrict our comparison to the region $-t<0.2$ GeV$^2$, where the incoherent contribution is expected to be small.
In addition, HERMES used a positron beam rather than an electron beam. Since the DVCS-BH interference term changes sign with the lepton charge, the measured beam-spin asymmetry also reverses sign. To facilitate a direct comparison with our calculations, which assume an electron beam, we therefore compare against $-A_{LU}^{\sin \phi}$ for the HERMES results.
Overall, the calculations are broadly consistent with the measured data.

\section{Physics Projection for Polarized \He3 at the EIC}

At the EIC, a wide variety of ion beams with be available for nuclear physics studies~\cite{Abdul_Khalek_2022, Atoian:2025dib}, including \He3, which serves as an effective source of highly polarized neutrons.
For a spin-$1/2$ nucleus such as \He3, the nuclear spin gives rise to additional asymmetries beyond the lepton asymmetry $A_{LU}$. These observables provide sensitivity to the polarized CFFs $\tilde {\mathcal H}$ and $\tilde {\mathcal E}$, which are probed by longitudinal and transverse nuclear polarization, respectively.
The simplest nuclear-spin observable, which we investigate in this section, is the nuclear longitudinal single-spin asymmetry,
\begin{equation}
    A_{UL}=\frac{1}{P_h}\frac{d\sigma^\Rightarrow - d\sigma^\Leftarrow}{d\sigma^\Rightarrow + d\sigma^\Leftarrow}\,,
\end{equation}
where $d\sigma^{\Rightarrow(\Leftarrow)}$ is the differential cross section for a target nucleus polarized parallel (antiparallel) to the lepton beam, and $P_h$ is the degree of nuclear polarization.
In the limit of Bethe--Heitler dominance and in EIC kinematics, the leading terms of this asymmetry follow:
\begin{equation}
\begin{aligned}
    A_{UL}
    &\sim \frac{x_A(1+\epsilon^2)^2}{y}
    \frac{s_{1,LP}^{\mathcal I}\sin \phi + \cdots}
    {c_0^{BH}+\cdots} \\
    &\propto \left(
    F_1 \IM \tilde{\mathcal H}(\xi_A,t)
    + \xi_A(F_1+F_2)\IM \mathcal H(\xi_A,t)
    \right)\sin\phi \,,
\end{aligned}
\label{eq:AUL_lin}
\end{equation}
where $s_{1,LP}^{\mathcal I}$ denotes the interference $\sin\phi$-coefficient in the presence of longitudinal polarization of the target nucleus.
In addition to the longitudinal beam-spin and target-spin asymmetries discussed above, one may also define the longitudinal double-spin asymmetry,
\begin{equation}
    A_{LL}=\frac{1}{P_l P_h}\frac{d\sigma^{\rightarrow \Rightarrow} + d\sigma^{\leftarrow\Leftarrow} - d\sigma^{\rightarrow \Leftarrow} - d\sigma^{\leftarrow\Rightarrow}}{d\sigma^{\rightarrow \Rightarrow} + d\sigma^{\leftarrow\Leftarrow} + d\sigma^{\rightarrow \Leftarrow} + d\sigma^{\leftarrow\Rightarrow}}\,,
\end{equation}
which is sensitive to the real part of $\tilde {\mathcal H}(\xi_A ,t)$ ($\RE \tilde {\mathcal H}(\xi_A ,t)$). Additional observables involving transverse nuclear asymmetries provide access to the helicity-flip CFFs $ {\mathcal E}(\xi_A ,t)$ and $\tilde {\mathcal E}(\xi_A ,t)$. 
These asymmetries are beyond the scope of the present work and are therefore deferred to future studies.
This work assumes that the contribution of ${\mathcal E}$ and $\tilde {\mathcal E}$ are negligible in the observables discussed here at EIC kinematics; see the Supplementary Material for more detail.

\begin{figure}
    \centering
    \includegraphics[width=1\linewidth]{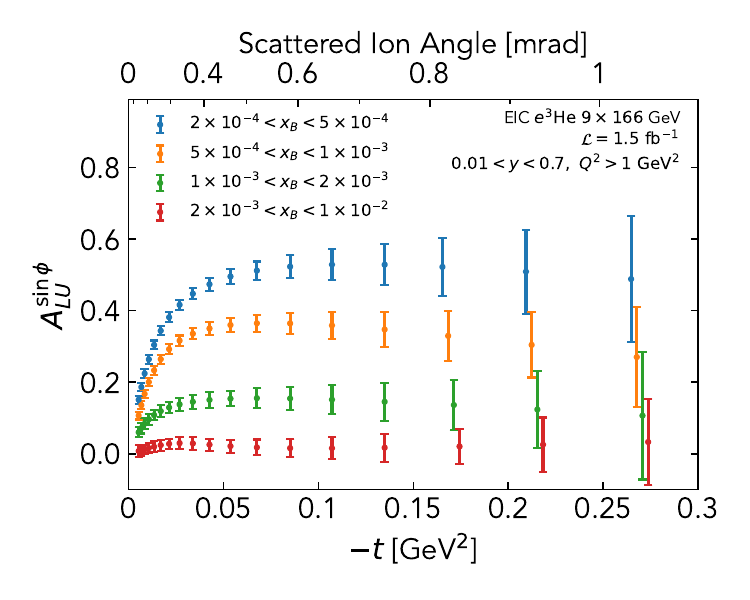}
    \caption{Projected electron-spin asymmetry $\sin \phi$-coefficient $A_{LU}^{\sin \phi}$ for $e^3$He collisions at $9\times166$ GeV, presented as a function of $-t$ in several bins of $x_B$.
    Uncertainties are the projected statistical uncertainties assuming 1.5 fb$^{-1}$ of luminosity and electron polarization of $P_e=0.7$.
    The upper axis indicates the ion scattering angle corresponding with each value of $-t$.
    Here and in Figs.~\ref{fig:ImH_EIC} and ~\ref{fig:ImHt_EIC} projected points have been shifted horizontally when overlapping.}
    \label{fig:ALU_EIC}
\end{figure}

\begin{figure}
    \centering
    \includegraphics[width=1\linewidth]{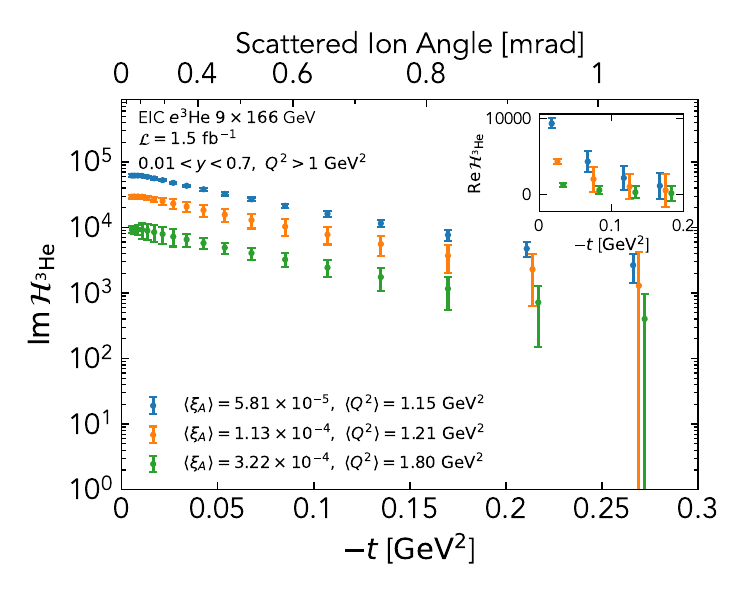}
    \caption{\textit{Main figure:} Projected $\IM \mathcal H$ extracted from $A_{LU}$ measurements using $e^3$He collisions at $9\times166$ GeV, presented as a function of $-t$ at several values of skewness $\xi_A$.
    Uncertainties are the projected statistical uncertainties assuming 1.5 fb$^{-1}$ of luminosity and electron polarization of $P_e=0.7$.
    The upper axis indicates the ion scattering angle corresponding with each value of $-t$.
    Here the largest-$x_B$ bins from Fig.~\ref{fig:ALU_EIC} have been merged.
    \textit{Inset:} Projected $\RE \mathcal H$ using the same settings and luminosity.}
    \label{fig:ImH_EIC}
\end{figure}

\begin{figure}
    \centering
    \includegraphics[width=1\linewidth]{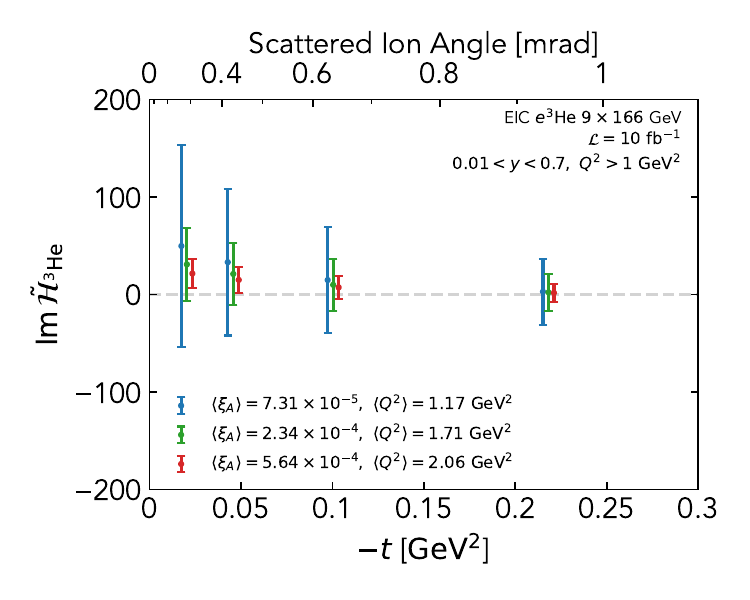}
    \caption{Projected $\IM \Ht$ extracted from $A_{UL}$ measurements using $e^3$He collisions at $9\times166$ GeV, presented as a function of $-t$ at several values of skewness $\xi_A$.
    Uncertainties are the projected statistical uncertainties assuming 10 fb$^{-1}$ of luminosity and ion polarization of $P_h=0.7$.
    A dashed grey line is shown at $\IM \Ht=0$ for reference.
    Here the smallest-$x_B$ bins from Fig.~\ref{fig:ALU_EIC} have been merged.
    The upper axis indicates the ion scattering angle corresponding with each value of $-t$.}
    \label{fig:ImHt_EIC}
\end{figure}

\begin{figure}
    \centering
    \includegraphics[width=1\linewidth]{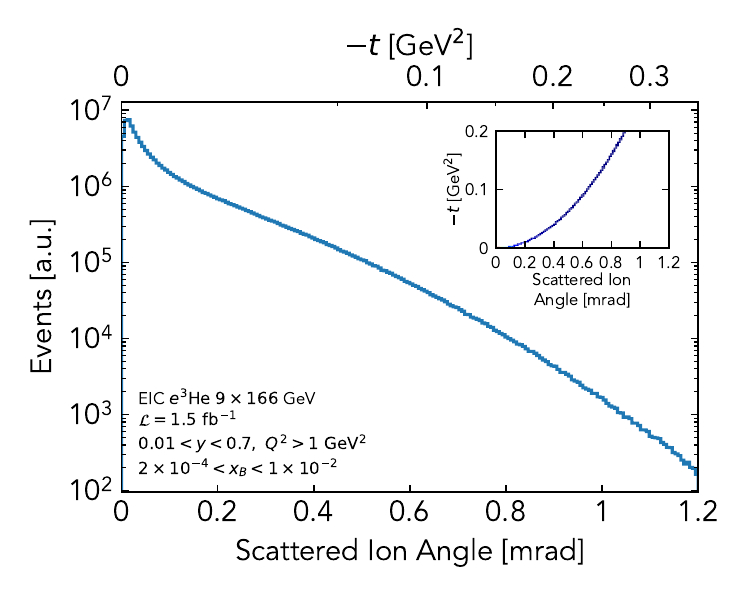}
    \caption{\textit{Main figure:} Angular distribution of the final-state $^3$He ion in  $e^3\text{He}\rightarrow e^3\text{He}\gamma$ at $9\times166$ GeV, integrated over all selected kinematics in the studied region $2\times10^{-4} < x_B < 1\times 10^{-2}$.
    The upper axis shows the value of $-t$ corresponding to the ion scattering angle.
    \textit{Inset:} Two-dimensional distribution of the scattered ion angle and $-t$ for selected events.
    }
    \label{fig:ion_kinematics}
\end{figure}

In order to assess the sensitivity of the EIC to this process, we generate events corresponding to the anticipated beam configuration during the early stages of EIC operation, consisting of a $9$-GeV electron beam colliding with a $166$-GeV/nucleon \He3 beam. Consistent with the current EIC accelerator design, both the electron and ion beam polarizations are assumed to be $P_e=P_h=0.7$.
Events are generated within the kinematic range $0.01<y<0.7$, $0<-t<0.5$ GeV$^2$, and $Q^2>1$ GeV$^2$.
Final-state electrons and photons are required to fall within the central ePIC detector acceptance, defined by $|\eta|<3.5$ and $p_T>0.2$ GeV$/c$.
The scattered $^3$He ions are required to satisfy $\theta <5$ mrad with respect to the beamline, matching the acceptance of the Roman Pot detectors in the ePIC far-forward region \cite{Abdul_Khalek_2022}.
We discuss later the impact of the minimum angular acceptance for the far-forward detectors, which will be driven by beam conditions but is yet undetermined; to aid in understanding the impact of acceptance on Figs.~\ref{fig:ALU_EIC},~\ref{fig:ImH_EIC}, and~\ref{fig:ImHt_EIC}, the ion scattering angle is displayed as an alternate axis, as it correlates strongly with $t$ (see the inset of Fig.~\ref{fig:ion_kinematics}).
Events passing these selection criteria are weighted according to the total cross section and the assumed integrated luminosity. The resulting event yields are then used to estimate rates in individual kinematic bins and to evaluate the projected sensitivity to asymmetries and CFFs using the method of Fisher information~\cite{Cowan:1998ji,Barlow}; see the Supplementary Materials for more detail.

Fig.~\ref{fig:ALU_EIC} shows the projected measurement of the electron-spin asymmetry $A^{\sin \phi}_{LU}$ as a function of $-t$ in several bins of $x_B$, assuming the early-science integrated luminosity of 1.5 fb$^{-1}$ divided evenly between positive- and negative-helicity electron-beam configurations.
Event yields are calculated in bins of $x_B$, $t$, and $\phi$, and the $\phi$-dependent fit of Eq.~\ref{eq:fit} is used to propagate the statistical uncertainties to the extracted coefficient $A_{LU}^{\sin\phi}$.
The predicted electron-spin asymmetry increases markedly with decreasing $x_B$, reflecting the growth of $\IM \mathcal H$ as parton density increases.
The coherent Bethe--Heitler and DVCS event rates are expected to be sufficiently large to enable a precise measurement of the $t$ dependence of $A_{LU}^{\sin\phi}$ up to approximately $-t \simeq 0.2~\mathrm{GeV}^2$, beyond which the coherent cross section falls rapidly.
We also observe that, at small $x_B$, the growth of both $\IM\mathcal H$ and $\RE\mathcal H$ causes the DVCS contribution to become increasingly important relative to the Bethe--Heitler process. Consequently, the total cross section is no longer strongly dominated by Bethe--Heitler, and the linearized approximations for the beam- and target-spin asymmetries cease to provide an accurate description.

To determine the statistical precision with which $\IM \mathcal H$ and $\RE \mathcal H$ can be extracted in this nonlinear regime, we employ a more complete description of the $\phi$-dependence of $A_{LU}$ while working under the assumption that the CFF $\mathcal H$ dominates the DVCS amplitudes within our kinematics.
We follow a similar approach to Ref.~\cite{Hattawy_2017} to extract both components of $\HA$ from the asymmetry, but modify the coefficient functions as appropriate for a spin-1/2 nucleus, parameterizing the asymmetry as
\begin{equation}
\begin{aligned}
    &A_{LU}(\phi)\\
    &=\frac{\alpha_0(\phi) \IM \mathcal H}{\alpha_1(\phi) + \alpha_2(\phi) \RE \mathcal H + \alpha_3(\phi)\left(\left(\RE \mathcal H\right)^2 + \left(\IM \mathcal H\right)^2\right)}\,,
\end{aligned}
\end{equation}
with
\begin{align}
    &\alpha_0(\phi) \IM \mathcal H = \mathcal K_{\mathcal I} \left(s_1^{\mathcal I}\sin \phi+\cdots\right)\,, \\
    &\alpha_1(\phi) = \mathcal K_{BH}\left(c_0^{BH} + c_1^{BH} \cos \phi + c_2^{BH} \cos 2\phi + \cdots\right)\,,\\
    &\alpha_2(\phi) \RE \mathcal H = \mathcal K_{\mathcal I} \left(c_0^{\mathcal I} + c_1^{\mathcal I}\cos \phi+\cdots\right)\,,\\
    &\alpha_3(\phi)\left(\left(\RE \mathcal H\right)^2  + \left(\IM \mathcal H\right)^2\right) = \mathcal K_{DVCS}\left(c_0^{DVCS} + \cdots\right)\,,
\end{align}
where the harmonic coefficients for a spin-1/2 hadron are taken from Ref.~\cite{Belitsky_2002}.
For a spin-1/2 nucleus such as \He3, additional contributions involving $\Ht$, ${\mathcal E}$, and $\tilde{\mathcal E}$ are present in the full expression. 
However, these terms are suppressed by powers of $x_B$ and are therefore neglected in the present analysis; see the Supplementary Material for a more detailed discussion.
This parameterization, together with the projected event yields in bins of $x_B$, $t$, and $\phi$, enables the propagation of statistical uncertainties on the measured yields to the extracted values of $\IM \mathcal H$ and $\RE \mathcal H$. 
While $\IM\mathcal H$ is primarily constrained by the odd harmonic components of $A_{LU}$, $\RE\mathcal H$ is constrained more weakly through the even harmonic terms appearing in the denominator.
Fig.~\ref{fig:ImH_EIC} shows the projected measurements of $\IM \mathcal H$ as a function of $t$ for several values of the skewness $\xi_A$, along with an inset displaying the corresponding projections for $\RE \mathcal H$.
We predict that an integrated luminosity of 1.5 fb$^{-1}$ will enable precise differential measurements of $\IM \mathcal H$ up to $-t\sim0.2$ GeV$^2$ at several values of skewness $\xi_A$. The corresponding constraints on $\RE \mathcal H$ are expected to be less precise, although additional sensitivity may be obtained through measurements of the absolute cross section. Such measurements, however, are experimentally more challenging and are not considered further in this study.

The ion-spin asymmetry $A_{UL}$ and the resulting sensitivity to $\IM \Ht$ are also examined.
The event yields as a function of $x_B$, $t$, and $\phi$ are used to determine the statistical uncertainties on the extracted value of $\IM \Ht$ in the same manner as for $\IM \mathcal H$, using the following functional form:
\begin{equation}
\begin{aligned}
    &A_{UL}(\phi)=\\
    &\frac{\tilde \alpha_0(\phi) \left( \IM \Ht +\xi\frac{F_1+F_2}{F_1}\IM \HA\right)}{\alpha_1(\phi) + \alpha_2(\phi) \RE \mathcal H + \alpha_3(\phi)\left(\left(\RE \mathcal H\right)^2 + \left(\IM \mathcal H\right)^2\right)}\,,
\end{aligned}
\end{equation}
Motivated by the substantially smaller asymmetry signal, we assume an integrated luminosity of 10 fb$^{-1}$, divided equally between the two nuclear-polarization settings configurations.
Although the $\phi$-dependence of $A_{UL}$ is sensitive to $\IM \mathcal H$ and $\RE \mathcal H$ in addition to $\IM \Ht$, we assume that the components of $\mathcal H$ have been constrained independently through measurements of $A_{LU}$.
Under this assumption, the statistical uncertainties on $A_{UL}$ dominate the extraction of $\IM \Ht$.
Fig.~\ref{fig:ImHt_EIC} shows the projected measurements of $\IM \tilde{\mathcal H}$ as a function of $t$ for several values of the skewness $\xi_A$.
The projected precision on $\IM \Ht$ is significantly worse than that obtained for either $\IM \mathcal H$ or $\RE \mathcal H$, despite the larger integrated luminosity assumed in this study.
This reduced sensitivity arises from the relatively small magnitude of $\IM\Ht$ compared with $\IM\mathcal H$, which leads to a substantially weaker target-spin asymmetry than the corresponding electron-spin asymmetry.
With an integrated luminosity of 10 fb$^{-1}$, the projected measurements provide meaningful constraints on the $\xi_A$ and $t$ dependence of $\IM \Ht$, particularly at larger values of $x_B$ and $\xi_A$.
More precise measurements of $\IM \Ht$ would require substantially larger data sets of $e^3$He to be collected at EIC.

The isolation of DVCS measurement is most readily possible when exclusively reconstructing all final-state particles in the event, including detecting the scattered ion.
Such exclusivity is important for suppressing the substantial background from $\pi^0$ production, and in the case of nuclear DVCS, for separating the coherent process from incoherent nuclear breakup.
To access the feasibility of this measurement, we examine the capability of the EIC far-forward detector system to detect the scattered $^3$He ions.
Fig.~\ref{fig:ion_kinematics} presents the angular distribution of the recoil $^3$He ions in simulated $e^3\text{He}\rightarrow e^3\text{He}\gamma$ events, together with the correlation between this scattering angle and the momentum transfer $t$.
The predicted scattering angles lie well within the nominal outer acceptance of the Roman Pot and off-momentum detectors. However, a substantial fraction of the recoil ions are nearly collinear with the beam.
Because the longitudinal momentum-transfer $\xi_A$ is small at EIC energies, the total momentum-transfer of the reaction is largely transverse ($|t| - p_T^2<10^{-4}\text{ GeV}^2$ within our chosen bins), resulting in a strong correlation between the ion scattering angle and $t$.
At EIC energies, the longitudinal momentum transfer is very small, so that the momentum transfer is predominantly transverse, with $|t|-p_T^2<10^{-4}~\mathrm{GeV}^2$ throughout the kinematic region considered in this work. Consequently, the recoil-ion scattering angle is strongly correlated with $|t|$, implying that the minimum-angle acceptance of the far-forward detectors will determine the minimum accessible value of $|t|$ for fully exclusive nuclear DVCS measurements.

The majority of scattered ions in coherent DVCS are produced at $-t<0.25$ GeV$^2$ and at scattering angles below 1~mrad, placing stringent requirements on the minimum angle acceptance of the EIC far-forward detectors. Although the achievable inner acceptance has not yet been established for $e$+\He3 running, the Roman Pot detectors, which provide the relevant capability for tagging ions with small deflections, will ultimately be limited by the transverse extent and divergence of the ion beam. As a reference, experience from HERA indicates that detector operation at distances of approximately $10\sigma$ from the beam may be achievable~\cite{osti_1765663}. The corresponding beam divergence and minimum approach distance have not yet been established for \He3 at the EIC, and a substantial fraction of coherent DVCS events may therefore fall outside the Roman Pot acceptance under standard high-luminosity optics.

This limitation is particularly relevant for nuclear imaging. With a \He3 nuclear radius of $R_A\approx2~\mathrm{fm}$~\cite{doi:10.1126/science.adj2610}, the characteristic momentum scale is $-t\sim0.01$ GeV$^2$, well below that accessible with a minimum scattering angle of order $1~\mathrm{mrad}$. Extending the acceptance toward smaller deflection angles will therefore likely require dedicated beam-optics configurations, as discussed in Ref.~\cite{y4yv-y9dn}. A large-acceptance configuration with reduced beam divergence near the Roman Pots could potentially extend ion tagging to $-t\sim0.05~\mathrm{GeV}^2$, albeit at the expense of instantaneous $e$+\He3 luminosity. Reaching the characteristic nuclear-imaging region of $-t\sim0.01~\mathrm{GeV}^2$ and below may require more substantial modifications to the interaction-region optics, such as an additional beam focus \cite{y4yv-y9dn}. Such capabilities could therefore provide an important consideration for the design of a potential second EIC interaction region and detector.

At very small $|t|$, the enhancement of coherent DVCS relative to incoherent backgrounds may allow partial recovery of events even in the absence of direct recoil-ion tagging. Such measurements, however, will require careful control of residual backgrounds and efficient veto capabilities. Extending the far-forward acceptance to the smallest possible recoil-ion scattering angles therefore remains important for establishing event exclusivity and maximizing the accessible $t$ range. This capability is relevant not only for coherent nuclear DVCS, but also for a broader program of exclusive nuclear measurements at the EIC, including Deeply Virtual Meson Production \cite{PhysRevC.99.015203, wpbg-h511, KESLER2026140585} and the photoproduction of heavy vector mesons \cite{Chang:2021jnu, Gryniuk:2020mlh}.


\section{Conclusion}

We have presented a model for coherent DVCS on polarized \He3 and applied it to simulations of the process at the EIC.
Projections for measurements of the electron-spin asymmetry $A_{LU}$
from \He3, as well as for the extraction of CFF components $\IM \mathcal H_{^3\text{He}}$, $\RE \mathcal H_{^3\text{He}}$, and $\IM \Ht_{^3\text{He}}$, have been performed using this framework.
Our results indicate that early $e$\He3 data will enable precise differential measurements of $\IM \mathcal H_{^3\text{He}}$ and provide significant constraints on $\RE\mathcal H_{^3\text{He}}$.
By contrast, meaningful constraints on $\IM \Ht_{^3\text{He}}$ will require substantially larger integrated luminosities, particularly for access to small $\xi_A$.
The kinematics of the scattered \He3 ion in this process has been examined as well. 
The exclusivity requirement of the coherent nuclear process places stringent demands on the minimal-angle acceptance of the EIC far-forward detectors in order to enable precision measurements of coherent nuclear DVCS. 
Nonetheless, the EIC can be expected to provide unique and novel constraints on the three-dimensional partonic structure of polarized nuclei into the kinematics of low $x_B$.

Looking ahead, the present framework can be incorporated into a full ePIC detector simulation to quantify detector acceptance, reconstruction efficiencies, and systematic effects under realistic experimental conditions. 
The model may also be extended to include transverse-spin observables, providing access to the helicity-flip CFFs $\mathcal E_{^3\mathrm{He}}$ and $\tilde{\mathcal E}_{^3\mathrm{He}}$ and enabling a more complete investigation of the spin structure of \He3. 
An equally important direction is the development of a corresponding description of incoherent nuclear DVCS. 
In addition to constituting a background to coherent scattering, the incoherent process provides direct sensitivity to the structure of bound nucleons and offers unique opportunities for polarized-neutron tomography necessary for full flavor decomposition of the nucleon. 
Together, coherent and incoherent DVCS on polarized \He3 will provide complementary probes of nuclear and nucleonic structure, forming a unified program for three-dimensional imaging at the EIC.

\section*{Acknowledgements}
The authors would like to thank John Terry, Ivan Vitev, Duff Neill and Jen-Chieh Peng for the valuable discussions and insightful feedback. This work was supported by the U.S. Department of Energy through the Los Alamos National Laboratory. Los Alamos National Laboratory is operated by Triad National Security, LLC, for the National Nuclear Security Administration of U.S. Department of Energy (Contract No. 89233218CNA000001). Research presented in this article was supported by the Laboratory Directed Research and Development program of Los Alamos National Laboratory under project number 20260377ER.

\bibliographystyle{apsrev4-2}
\bibliography{references}

\end{document}